\documentclass[leqno]{article}
\usepackage{bm}
\usepackage{ascmac}
\usepackage{amsmath}
\usepackage{amssymb}
\usepackage{amsfonts}
\usepackage{graphicx}
\usepackage{slashbox}
\usepackage{subfigure}
\usepackage{multirow}

\usepackage{jjss-2e}

\makeatletter

\@addtoreset{equation}{section}
\makeatother
\allowdisplaybreaks

\frompage{0}
\title{Bayesian index of superiority and the $p$-value of the conditional test for Poisson parameters}
\shorttitle{Bayesian Index and $p$-value for Poisson Parameters}
\author{ 
Masaaki Doi
\footnote{
Clinical Data Science and Quality Management Department, Toray Industries, Inc., 1-1, Nihonbashi-muromachi 2-chome, Chuo-ku, Tokyo 103-8666, Japan.
}
\footnote{
Graduate School of Science and Engineering, Chuo University, 1-13-27, Kasuga, Bunkyo-ku, Tokyo 112-8551, Japan.
}
} 
\abstract{ 
We consider the problem of comparing two Poisson parameters from the Bayesian perspective. Kawasaki and Miyaoka (2012b) proposed the Bayesian index $P(\lambda_1 < \lambda_2\, |\, X_1,X_2)$ and expressed it using the hypergeometric series. In this paper, under some conditions, we give four other expressions of the Bayesian index in terms of the cumulative distribution functions of beta, $F$, binomial, and negative binomial distribution. Next, we investigate the relationship between the Bayesian index and the $p$-value of the conditional test with the null hypothesis $H_0: \lambda_1 \geq \lambda_2 $ versus an alternative hypothesis $H_1: \lambda_1<\lambda_2 $. Additionally, we investigate the generalized relationship between $P\left( \lambda_1/\lambda_2 <c\, |\, X_1,X_2 \right)$ and the $p$-value of the conditional test with the null hypothesis $H_0:\lambda_1/\lambda_2 \geq c$ versus the alternative $H_1: \lambda_1/\lambda_2 < c$. We illustrate the utility of the Bayesian index using  analyses of real data. Our finding suggests that $\theta$ can potentially be useful in an epidemiology and in a clinical trial.}
\keywords{
Bayesian index, beta distribution, binomial distribution, conditional power prior, conditional test, non-informative prior, Poisson parameters.
}
\begin{document}
%
%
\maketitle

\section{Introduction}
Comparing two groups is one of the most popular topics in statistics. For comparison,  the frequentist methods are often applied in medical statistics and epidemiology. However, in recent years, Bayesian methods have gained increasing attention 
because prior information can be used to improve the efficiency of inference. Particularly for  categorical data analysis, to evaluate the superiority of one group to  another from the Bayesian perspective, 
Kawasaki and Miyaoka (2012a; 2012b; 2014), Kawasaki et al. (2013; 2014),
Altham (1969), and Howard (1998) investigated the posterior probability
$P(\theta_1 > \theta_2 \,|\, X_1,X_2)$ where $X_1, X_2$ are data
and $\theta_1, \theta_2$ are parameters of interest in each group.
For binomial distributions with proportions $\pi_1,\pi_2$, Kawasaki and Miyaoka (2012a) named $\theta=P(\pi_{1} > \pi_{2}\,|\, X_1,X_2)$ a Bayesian index and expressed it by the hypergeometric series. Moreover, Kawasaki et al. (2014) showed that $\theta$ and the one sided $p$-value of Fisher's exact test are equivalent under certain conditions. A similar relationship is investigated by Altham (1969) and Howard (1998).

For Poisson distributions with parameters $\lambda_1,\lambda_2$, Kawasaki and Miyaoka (2012b) proposed a Bayesian index $\theta=P(\lambda_{1} < \lambda_{2}\,|\, X_1,X_2)$, expressed it using the hypergeometric series, and inferred the relationship between $\theta$ and the one-sided $p$-value of the z-type Wald test. However, hypergeometric series are, in general, difficult to calculate, and the exact relationship between $\theta$ and $p$-value was not established. In this paper, we give other expressions of the Bayesian index, which can be easily calculated, and show the exact relationship between $\theta$ with non-informative prior and the one-sided $p$-value of the conditional test. Additionally, we investigate the relationship between the generalized version of the Bayesian index and the $p$-value of the conditional test with more general hypotheses.

The remainder of this paper is structured as follows. In section 2, we give four expressions of the Bayesian index $P\left(\lambda_1 < \lambda_2\, |\, X_1,X_2 \right)$ other than hypergeometric series under some conditions. In section 3, we investigate the relationship between the Bayesian index and the $p$-value of the conditional test with the null hypothesis $H_0: \lambda_1 \geq \lambda_2 $ versus the alternative hypothesis $H_1: \lambda_1 < \lambda_2$. In section 4, as a generalization, we investigate the relationship between $P\left(\lambda_1/\lambda_2 < c \right)$ and the $p$-value of the conditional test with the null hypothesis $H_0: \lambda_1/\lambda_2 \geq c $ versus the alternative $H_1: \lambda_1/\lambda_2 < c$. 
In section 5, we illustrate the Bayesian index using analyses of real data. Finally, we provide some concluding remarks in section 6.
\section{Bayesian index for the Poisson parameters and its expressions}
\subsection{Bayesian index with the gamma prior}
We consider two situations. First, for $i=1,2$ and $j=1,\ldots, n_i$, let $X_{ij}$ be the outcome of $j$th subject in the $i$th group and independently follow the Poisson distribution $P(\lambda_i )$, and let $X_i = \sum_{j=1}^{n_i} X_{ij}$. Second, for $i=1,2$, let $X_i$ be the independent Poisson process with Poisson rate $\lambda_i$ and let $n_i$ be the person-years at risk. For both cases, $X_i \stackrel{ind}{\sim} Po(n_i \lambda_i)$. In the following, let $n_1,n_2$ be the fixed integers for simplicity.

For the Bayesian analysis, let the prior distributions of $\lambda_i$ be $Ga(\alpha_i,\beta_i)$ with $\alpha_i,\beta_i>0$ for $i=1,2$, whose probability density function is $f(\lambda_i\,|\, \alpha_i, \beta_i) = \beta_i^{\alpha_i}/\Gamma(\alpha_i) \cdot \lambda_i^{\alpha_i-1} \exp(-\beta_i \lambda_i) $. Let $X_i=k_i$, $a_i:= \alpha_i+k_i$, and $b_i:= \beta_i+n_i$, then the posterior distributions of $\lambda_i$ is $Ga(a_i, b_i )$. Here, if $\alpha_i \in \mathbb{N}$, then $a_i \in \mathbb{N}$ for $i=1,2$.  However, in the following, we suppose that $\alpha_1,\alpha_2,\beta_1,\beta_2>0$ and $k_1,k_2 \in \mathbb{N} \cup \{ 0 \}$.

When the posterior of $\lambda_i$ is $Ga(a_i,b_i)$ for $i=1,2$, Kawasaki and Miyaoka (2012b) proposed the Bayesian index $\theta=P(\lambda_1 < \lambda_2\, |\, X_1, X_2)$ and derived the following expression:
\begin{eqnarray}
\theta
&=& 1- \frac{1}{a_2 B(a_1,a_2) } \left(
\frac{b_2}{b_1+b_2}
\right)^{a_2} \cdot {}_2 F_{1} \left( a_2, 1-a_1; 1+a_2; \frac{b_2}{b_1+b_2} \right) \label{p_k}
\end{eqnarray}
where 
$${}_2 F_{1}( a, b ;c; z) = \sum\limits_{t=0}^{\infty} \frac{(a)_t (b)_t }{(c)_t} \cdot \frac{z^t}{t!}
\hspace{5mm}(|z| < 1)
$$ is the hypergeometric series and $(k)_t$ is the Pochhammer symbol, that is, $(k)_0=1$ and $(k)_t=k(k+1) \cdot (k+t-1)$ for $t \in \mathbb{N}$. Let $F_{\nu_1,\nu_2}(x)$ be the cumulative distribution function of $F$ distribution with degrees of freedom $(\nu_1,\nu_2)$, that is, 
\begin{eqnarray}
F_{\nu_1,\nu_2}(x) = \int_0^x \frac{1}{zB\left( \nu_1/2,\nu_2/2 \right)} \left(
\frac{\nu_1 z}{\nu_1z+\nu_2}
\right)^{\nu_1/2}\left(
\frac{\nu_2 }{\nu_1z+\nu_2}
\right)^{\nu_2/2} dz \label{F}
\end{eqnarray}
where $B(a,b)=\int_0^{\infty} x^{a-1} (1-x)^{b-1} dx$ is the beta function.
Then, we can obtain the following expressions of $\theta$.
\proclaimit{Theorem\n 1.}{
If the posterior distribution of $\lambda_i$ is $Ga(a_i,b_i)$ with $a_i, b_i>0$ for $i=1,2$, then the
Bayesian index $\theta=P( \lambda_1 < \lambda_2 \,|\, X_1,X_2)$ has the following two expressions:
\begin{eqnarray}
\theta &=& I_{\frac{b_1}{b_1+b_2}}(a_1,a_2) \label{theta1}\\[5pt]
&=& F_{2a_1,2a_2}\left(\frac{b_1/a_1}{b_2/a_2} \right) \label{theta2}
\end{eqnarray}
where 
\begin{eqnarray*}
I_x(a,b) = \frac{1}{B(a,b)} \int_0^x t^{a-1} (1-t)^{b-1} dt
\end{eqnarray*}
is the cumulative distribution function of the beta distribution, also known as the regularized incomplete beta function.
Moreover, if both $a_1$and $a_2$ are natural numbers, then $\theta$ has the following two additional expressions:
\begin{eqnarray}
\theta 
&=& \sum\limits_{r=0}^{a_2-1} {a_1+a_2-1 \choose r}
\left( \frac{b_2}{b_1+b_2} \right)^r 
\left( \frac{b_1}{b_1+b_2} \right)^{a_1+a_2-1-r} \label{theta3}\\[5pt]
&=& \sum\limits_{r=0}^{a_2-1} {a_1+r-1 \choose a_1 -1} \left( \frac{b_1}{b_1+b_2} \right)^{a_1} 
\left( \frac{b_2}{b_1+b_2} \right)^{r}. \label{theta4}
\end{eqnarray}
$(\ref{theta3})$ and $(\ref{theta4})$ are the cumulative distribution functions of the binomial and negative binomial distributions, respectively. 
}
{\sc Proof.}
First, (\ref{theta1}) can be shown by modifying (\ref{p_k}) using $I_{x}(a,b) = \frac{1}{B(a,b)} \cdot \frac{x^a}{a} \cdot {}_2F_1(a,1-b; 1+a; x) $ and $I_z(a,b)=1-I_{1-z}(b,a)$, which are 26.5.23 and 26.5.2 of Abramowitz and Stegun (1964), respectively. Next, (\ref{theta2}) can be shown by
changing variable $t=\nu_1z/(\nu_1z+\nu_2)$ for (\ref{F}) with $\nu_1=2a_1, \nu_2=2a_2$ and $x=(b_1/a_1)/(b_2/a_2)$. 

In the following, suppose that $a_1,a_2 \in \mathbb{N}$. (\ref{theta3}) can be shown by (\ref{theta1}), 26.5.2 of Abramowitz and Stegun (1964) above, and 
26.5.4 of Abramowitz and Stegun (1964): $\sum\limits_{r=a}^n {n \choose r}p^r (1-p)^{n-r} =I_{p}(a,n-a+1)$ as follows
\begin{eqnarray*}
\theta &=& I_{\frac{b_1}{b_1+b_2}}(a_1,a_2) \hspace{5mm}(\because (\ref{theta1}))\\
&=& 1- I_{\frac{b_2}{b_1+b_2}}(a_2,a_1) \hspace{5mm}(\because 26.5.2 \text{ of Abramowitz and Stegun } (1964))\\
&=& 1 - \sum_{r=a_2}^{a_1+a_2-1} { a_1+a_2-1 \choose r} \left(\frac{b_2}{b_1+b_2} \right)^r 
\left(
\frac{b_1}{b_1+b_2}
\right)^{a_1+a_2-1-r}\\
&&\hspace{1cm}(\because 26.5.4 \text{ of Abramowitz and Stegun }(1964))\\[5pt]
&=&  \sum_{r=0}^{a_2-1} { a_1+a_2-1 \choose r}
\left(\frac{b_2}{b_1+b_2} \right)^r 
\left(
\frac{b_1}{b_1+b_2}
\right)^{a_1+a_2-1-r}.
\end{eqnarray*}

Finally, (\ref{theta4}) can be shown by 8.352-2 of Zwillinger (2014) :
\begin{eqnarray}
\int_{x}^{\infty} e^{-t} t^n dt = n! \cdot e^{-x} \sum_{m=0}^n \frac{x^m}{m!} \hspace{5mm}(n \in \mathbb{N}, x \in \mathbb{R})
 \label{zw1}
\end{eqnarray}
 as follows
\begin{eqnarray*}
\theta &=& P(\lambda_1 < \lambda_2 \,|\, X_1, X_2) \nonumber\\
&=& \int_0^{\infty} \left( \int_{\lambda_1}^{\infty} \frac{b_2^{a_2}}{\Gamma(a_2)} \lambda_2^{a_2-1} \exp(- b_2 \lambda_2 ) d\lambda_2 \right) \frac{b_1^{a_1}}{\Gamma(a_1)}
\lambda_1^{a_1-1} \exp(-b_1 \lambda_1) d\lambda_1. \label{thm1_1}\\
&=&\int_0^{\infty} \left( \int_{b_2 \lambda_1}^{\infty} \frac{b_2^{a_2}}{\Gamma(a_2)} \left( \frac{\pi_2}{b_2} \right)^{a_2-1} \exp(- \pi_2 ) \frac{d\pi_2}{b_2} \right) \frac{b_1^{a_1}}{\Gamma(a_1)}
\lambda_1^{a_1-1} \exp(-b_1 \lambda_1) d\lambda_1\\
&&\hspace{2cm} (\because \pi_2=b_2 \lambda_2
)\nonumber\\[5pt]
&=& \int_0^{\infty} \left( \sum_{r=0}^{a_2-1} \frac{(b_2\lambda_1)^r}{r!} \exp(-b_2\lambda_1) 
\right) \frac{b_1^{a_1}}{\Gamma(a_1)}
\lambda_1^{a_1-1} \exp(-b_1 \lambda_1) d\lambda_1 \hspace{5mm}(\because (\ref{zw1}))\\
&=& \sum_{r=0}^{a_2-1} \frac{b_1^{a_1} b_2^r}{\Gamma(a_1)r!} \int_0^{\infty} \lambda_1^{a_1+r-1} \exp(-(b_1+b_2) \lambda_1) d\lambda_1 \\
&=& \sum_{r=0}^{a_2-1} \frac{b_1^{a_1} b_2^r}{\Gamma(a_1)r!} \int_0^{\infty} \left( \frac{\pi_1}{b_1+b_2} \right)^{a_1+r-1} \exp(-\pi_1) \cdot \frac{d\pi_1}{b_1+b_2} \hspace{5mm} (\because \pi_1= (b_1+b_2)\lambda_1)\\
&=& \sum_{r=0}^{a_2-1} \frac{\Gamma(a_1+r)}{\Gamma(a_1)r!} \left( \frac{b_1}{b_1+b_2} \right)^{a_1} \left( \frac{b_2}{b_1+b_2} \right)^{r}\\ 
&=& \sum_{r=0}^{a_2-1} {a_1+r-1 \choose a_1-1} \left( \frac{b_1}{b_1+b_2} \right)^{a_1} \left( \frac{b_2}{b_1+b_2} \right)^{r}.
\end{eqnarray*}
We have completed the proof of theorem 1.
\begin{flushright}
$\square$
\end{flushright}

Kawasaki and Miyaoka (2012b) expressed $\theta$ using the hypergeometric series and computed it by summing the series. However, in general, it is difficult to calculate the hypergeometric series. Additionally, it is also difficult to understand the relationship between $\theta$ and other distributions. On the other hand, our expressions above have two advantages. First, we can calculate $\theta$ easily using the cumulative distribution functions of well-known distributions. Second, we can find the relationship between $\theta$ and some values represented by these cumulative distribution functions. 
Particularly, from expression (\ref{theta1}), we can easily show the relationship between $\theta$ and the $p$-value of the conditional test in section 3.

Here, we note an assumption for theorem 1. At the beginning of this section, we supposed the priors to be gamma distributions. However, for theorem 1, we only need the posteriors to be gamma. Therefore,  as long as the posteriors are gamma, we need not assume the priors to be gamma.
\subsection{Examples of the prior distribution} \label{eg}
In this section, we consider several examples of the prior of $\lambda_i$. All of the following examples are gamma distribution or the limit of the gamma distribution, and all the posteriors are gamma.
\proclaimit{Example\n 1.}{Non-informative prior;}
The non-informative prior distribution is $f_1(\lambda_i ) \propto \lambda_i^{-1}$. This is an improper prior but can be considered the limit of $Ga(\alpha_i,\beta_i)$ when $(\alpha_i,\beta_i) \to (0,0)$. Here, for $k_i>0$, the posterior is $Ga(k_i, n_i)$. Therefore, when $k_1,k_2>0$, that is, $k_1,k_2 \in \mathbb{N}$, theorem 1 states 
\begin{eqnarray*}
\theta &=& I_{\frac{n_1}{n_1+n_2}}(k_1,k_2)\\
&=& F_{2k_1,2k_2}\left(\displaystyle \frac{n_1/k_1}{n_2/k_2} \right)\\
&=& \sum\limits_{r=0}^{k_2-1} {k_1+k_2-1 \choose r}
\left( \frac{n_2}{n_1+n_2} \right)^r 
\left( \frac{n_1}{n_1+n_2} \right)^{k_1+k_2-1-r}\\[5pt]
&=& \sum\limits_{r=0}^{k_2-1} {k_1+r-1 \choose k_1 -1} \left( \frac{n_1}{n_1+n_2} \right)^{k_1} 
\left( \frac{n_2}{n_1+n_2} \right)^{r}.
\end{eqnarray*}
When $k_i=0$, the probability density function of the posterior is 
\begin{eqnarray*}
f(\lambda_i\,|\, X_i) &\propto & \lambda_i^{-1} \cdot \frac{\lambda_i^0}{0!} \exp(-\lambda_i)= \lambda_i^{-1} \exp(-\lambda_i).
\end{eqnarray*}
Hence, the posterior is improper and not a gamma distribution. Therefore, theorem 1 cannot be applied.

\proclaimit{Example\n 2.}{Jeffrey's prior;}
The Jeffrey's prior distribution (Jeffrey, 1946) is $f_2(\lambda_i ) \propto \lambda_i^{-1/2}$. This is also an improper prior but can be considered the limit of $Ga(1/2,\beta_i)$ when $\beta_i \to 0$. Here, the posterior is $Ga(k_i+1/2, n_i)$ for $k_i \geq 0$. Therefore, theorem 1 states
\begin{eqnarray*}
\theta &=& I_{\frac{n_1}{n_1+n_2}}\left(k_1+ \frac{1}{2} ,k_2 +\frac{1}{2} \right)\\
&=& F_{2k_1+1,2k_2+1}\left(\displaystyle \frac{n_1/(k_1+\frac{1}{2})}{n_2/(k_2+ \frac{1}{2})} \right).
\end{eqnarray*}
Because $k_i + 1/2 \not\in \mathbb{N}$ for any $k_i \in \mathbb{N} \cup \{0 \}$, we cannot have expression (\ref{theta3}) and (\ref{theta4}).
\proclaimit{Example\n 3.}{Conditional power prior;}
Let the historical data $x_i^0 \sim Po(m_i \lambda_i)$. 
Then the likelihood for $\lambda_i$ is
\begin{eqnarray*}
L(\lambda_i \,|\, x^0_{i}) = \frac{(m_i \lambda_i)^{x^0_{i} }}{x^0_{i}!} \exp(-m_i \lambda_i).
\end{eqnarray*}
Here, an example of the conditional power prior distribution (Ibrahim and Chen, 2000) is given as
\begin{eqnarray*}
f_3(\lambda_i ) \propto L(\lambda_i\,|\, x^0_{i})^{a_{i}} \cdot f_1(\lambda_i) 
\end{eqnarray*}
where $a_{i}$ is the fixed parameter such that $0 < a_i \leq 1$ and $f_1(\lambda_i) \propto \lambda_i^{-1}$.
Thus
\begin{eqnarray*}
f_3(\lambda_i ) &\propto& \left( \lambda_i^{x^0_i} \exp(-m_i\lambda_i) \right)^{a_{i}} \cdot \lambda_i^{-1}\\
& = & \lambda_i^{a_{i}x^0_{i}-1}\exp(-a_{i}m_i \lambda_i).
\end{eqnarray*}
Hence, the prior of $\lambda_i$ is $Ga(a_{i}x^0_{i},a_{i}m_i )$ when $x_i^0>0$. The posterior is $Ga(a_{i}x^0_{i}+k_i ,a_{i}m_i+n_i )$. Therefore, theorem 1 states 
\begin{eqnarray*}
\theta &=& I_{\frac{a_{1}m_1+n_1}{(a_{1}m_1 + n_1)+(a_{2}m_2 +n_2)}}(a_{1}x_1^0 + k_1, a_{2}x_2^0+ k_2)\\
&=& F_{2(a_{1}x_1^0+k_1),2(a_{2}x_2^0+k_2)}\left(\displaystyle \frac{(a_{1}m_{1} +n_1)/(a_{1}x^0_{1}+k_1)}{(a_{2}m_{2}+n_2)/(a_{2}x^0_{2}+k_2)} \right).
\end{eqnarray*}
Here, because  $a_{1}x_{1}^0, a_{2}x_{2}^0 \not\in \mathbb{N}$ in general, we cannot have expression (\ref{theta3}) and (\ref{theta4}) in general. On the other hand, when $a_{1}x_{1}^0, a_{2}x_{2}^0 \in \mathbb{N}$, we have expression (\ref{theta3}) and (\ref{theta4}) as follows
\begin{eqnarray*}
\theta 
&=& \sum\limits_{r=0}^{a_2x_2^0+k_2-1} {
a_1x_1^0+k_1+a_2x_2^0+k_2
-1 \choose r} \\
&&\hspace{5mm} \times \left( \frac{a_2m_2+n_2}{a_1m_1+n_1+ a_2m_2+n_2 } \right)^r 
\left( \frac{a_1m_1+n_1}{a_1m_1+n_1+ a_2m_2+n_2 } \right)^{a_1x_1^0+k_1+a_2x_2^0+k_2-1-r}\\[5pt]
&=& \sum\limits_{r=0}^{a_2x_2^0+k_2-1} {
a_1x_1^0+k_1+r-1
 \choose a_1x_1^0+k_1-1} \\
&&\hspace{1.5cm} \times 
\left( \frac{a_1m_1+n_1}{a_1m_1+n_1+ a_2m_2+n_2 } \right)^{a_1x_1^0+k_1}
\left( \frac{a_2m_2+n_2}{a_1m_1+n_1+ a_2m_2+n_2 } \right)^r .
\end{eqnarray*}
\section{The relationship between the Bayesian index and the $p$-value of the conditional test}
For binomial proportions, Kawasaki et al. (2014) and Altham (1969) showed the relationship between the Bayesian index and the one-sided $p$-value of the Fisher's exact test under certain conditions. For Poisson parameters, a similar relationship holds between the Bayesian index and the one-sided $p$-value of the conditional test.
\subsection{Conditional test}
From the frequentist perspective, we consider the conditional test based on the conditional distribution of $X_1$ given $X_1+X_2=k_1+k_2$ (Przyborowski and Wilenski, 1940; Krishnamoorthy and Thomson, 2004). The probability function is
\begin{eqnarray*}
f(X_1=k_1 \, |\, X_1+X_2= k_1+k_2) \hspace{6.5cm}\\
\hspace{2cm} = \displaystyle {k_1+k_2 \choose k_1} \left(
\frac{n_1\lambda_1}{n_1\lambda_1+n_2\lambda_2}
\right)^{k_1} \left(
\frac{n_2\lambda_2}{n_1\lambda_1+n_2\lambda_2}
\right)^{(k_1+k_2)-k_1}.
\end{eqnarray*}
To test the null hypothesis $H_0: \lambda_1 \geq \lambda_2$ versus the alternative $H_1: \lambda_1 < \lambda_2 $, the $p$-value is
\begin{eqnarray}
p&=&P(X_1 \leq k_1 \, | \, X_1+X_2=k_1+k_2, \lambda_1=\lambda_2 ) \nonumber \\
&=& \sum_{r=0}^{k_1} {k_1+k_2 \choose r} \left(
\frac{n_1}{n_1+n_2}
\right)^{r} 
\left(
\frac{n_2}{n_1+n_2}
\right)^{k_1+k_2-r}. \label{p_value1}
\end{eqnarray}
\proclaimit{Lemma\n 1.}{
If $k_2 >0$, then the one-sided $p$-value of the conditional test with $H_0:\lambda_1 \geq \lambda_2$ vs. $H_1: \lambda_1 < \lambda_2$ has the following expressions:
\begin{eqnarray*}
p&=& I_{\frac{n_2}{n_1+n_2}}(k_2,k_1+1)\\
&=& F_{2k_2,2(k_1+1)}\left( \frac{n_2/k_2}{ n_1/(k_1+1) } \right)\\
&=& \sum_{r=0}^{k_1}{ k_2 +r -1 \choose k_2 - 1} \left(
\frac{n_2}{n_1+n_2}
\right)^{k_2}
\left(
\frac{n_1}{n_1+n_2}
\right)^{r}\\
&=& \frac{1}{k_2 B(k_2,k_1+1)} \left( \frac{n_2}{n_1+n_2} \right)^{k_2} \cdot {}_2F_1\left(k_2,-k_1;1+k_2; \frac{n_2}{n_1+n_2} \right).
\end{eqnarray*}
If $k_2=0$, then $p=1$.
}
{\sc Proof.} For $k_2>0$, the proof is similar to that of theorem 1 with $\alpha_1,\alpha_2 \in \mathbb{N}$. For $k_2=0$, from (\ref{p_value1}),
\begin{eqnarray*}
p&=& \sum_{r=0}^{k_1} {k_1 \choose r} \left(
\frac{n_1}{n_1+n_2}
\right)^{r} 
\left(
\frac{n_2}{n_1+n_2}
\right)^{k_1-r}=1.
\end{eqnarray*}
\begin{flushright}
$\square$
\end{flushright}
\subsection{The relationship between the Bayesian index and the $p$-value of the conditional test}
\proclaimit{Theorem\n 2.}{
If $k_2>0$, then between $\theta=P(\lambda_1 < \lambda_2 \,|\, X_1,X_2)$ given $X_1=k_1+1$,  $X_2=k_2$ and the one-sided $p$-value of the conditional test with $H_0:\lambda_1 \geq \lambda_2$ vs. $H_1:\lambda_1< \lambda_2$ given $X_1=k_, X_2=k_2$,  the following relation holds
\begin{eqnarray*}
\lim\limits_{(\alpha_1,\alpha_2,\beta_1,\beta_2)\to (0,0,0,0)} \theta = 1- p.
\end{eqnarray*}
}

{\sc Proof.} 
From lemma 1, the $p$-value given $X_1=k_1, X_2=k_1$ is
\begin{eqnarray*}
p
&=& I_{\frac{n_2}{n_1+n_2}}(k_2,k_1+1).
\end{eqnarray*}
Therefore, from the relation $I_z(a,b)=1-I_{1-z}(b,a)$,
\begin{eqnarray}
1-p&=&  I_{\frac{n_1}{n_1+n_2}}(k_1+1,k_2). \label{p_value2}
\end{eqnarray}
Given $X_1=k_1+1, X_2=k_2$, on the other hand, $a_1=\alpha_1+k_1+1, a_2=\alpha_2+k_2, b_1=\beta_1+n_1, b_2=\beta_2+n_2$. Therefore, the Bayesian index is
\begin{eqnarray}
\theta &=& I_{\frac{b_1}{b_1+b_2}}(a_1,a_2) \nonumber \\
&=& I_{\frac{\beta_1+n_1}{\beta_1+n_1+ \beta_2+n_2}}(\alpha_1+k_1+1,\alpha_2+k_2). \label{theta10}
\end{eqnarray}
From (\ref{p_value2}) and (\ref{theta10}), 
\begin{eqnarray*}
\lim\limits_{(\alpha_1,\alpha_2,\beta_1,\beta_2)\to (0,0,0,0)} \theta = 1- p
\end{eqnarray*}
holds. We have just completed the proof of theorem 2.
\begin{flushright}
$\square$
\end{flushright}

Here, $\lim\limits_{(\alpha_1,\alpha_2,\beta_1,\beta_2)\to (0,0,0,0)} \theta$ equals the Bayesian index with non-informative priors when $k_1,k_2>0$.
\subsection{Plot of $\theta$ and $p$-value}
In this section, we plot and compare the Bayesian index $\theta=P(\lambda_1 < \lambda_2\,|\, X_1,X_2)$ with the non-informative prior and $1-p$ of the conditional test. First, we calculate $\theta$ and the one-sided $p$-value for all pairs of $(X_1,X_2)$ satisfying $1 \leq X_1 \leq 2n_1$ and $1 \leq X_2 \leq 2n_2$, for $n_1=n_2=10,20,50,$ and $100$, respectively. In Figure 1, the horizontal axis shows the difference between sample rates $\widehat{\lambda}_1 - \widehat{\lambda}_2$ where $\widehat{\lambda}_i = X_i/n_i$, and the vertical axis shows the difference $\theta - (1-p)$. We can see that $\theta$ is always greater than $1-p$. Moreover, $\theta - (1-p)$ tends to be greater when $|\widehat{\lambda}_1 - \widehat{\lambda}_2|$ is small, and $\theta - (1-p)$ tends to decrease as $n_1,n_2$ increases.
\begin{figure}[h]
\begin{center}
\begin{minipage}{100mm}
\subfigure[$n_1=n_2=10$ \label{figure1a}]{
\resizebox*{5cm}{!}{\includegraphics[height=5cm]{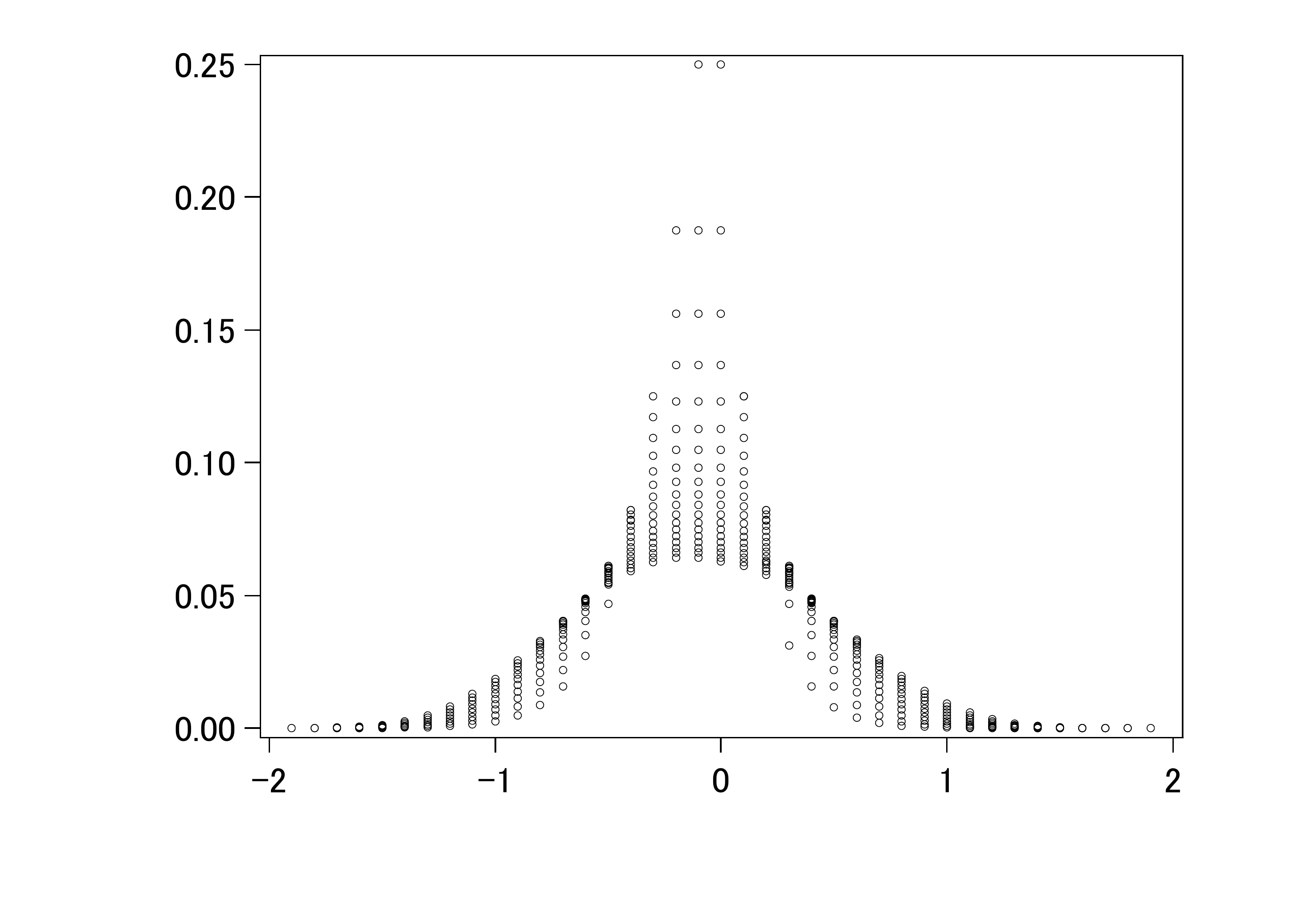}}}%
\subfigure[$n_1=n_2=20$ \label{figure1b}]{
\resizebox*{5cm}{!}{\includegraphics[height=5cm]{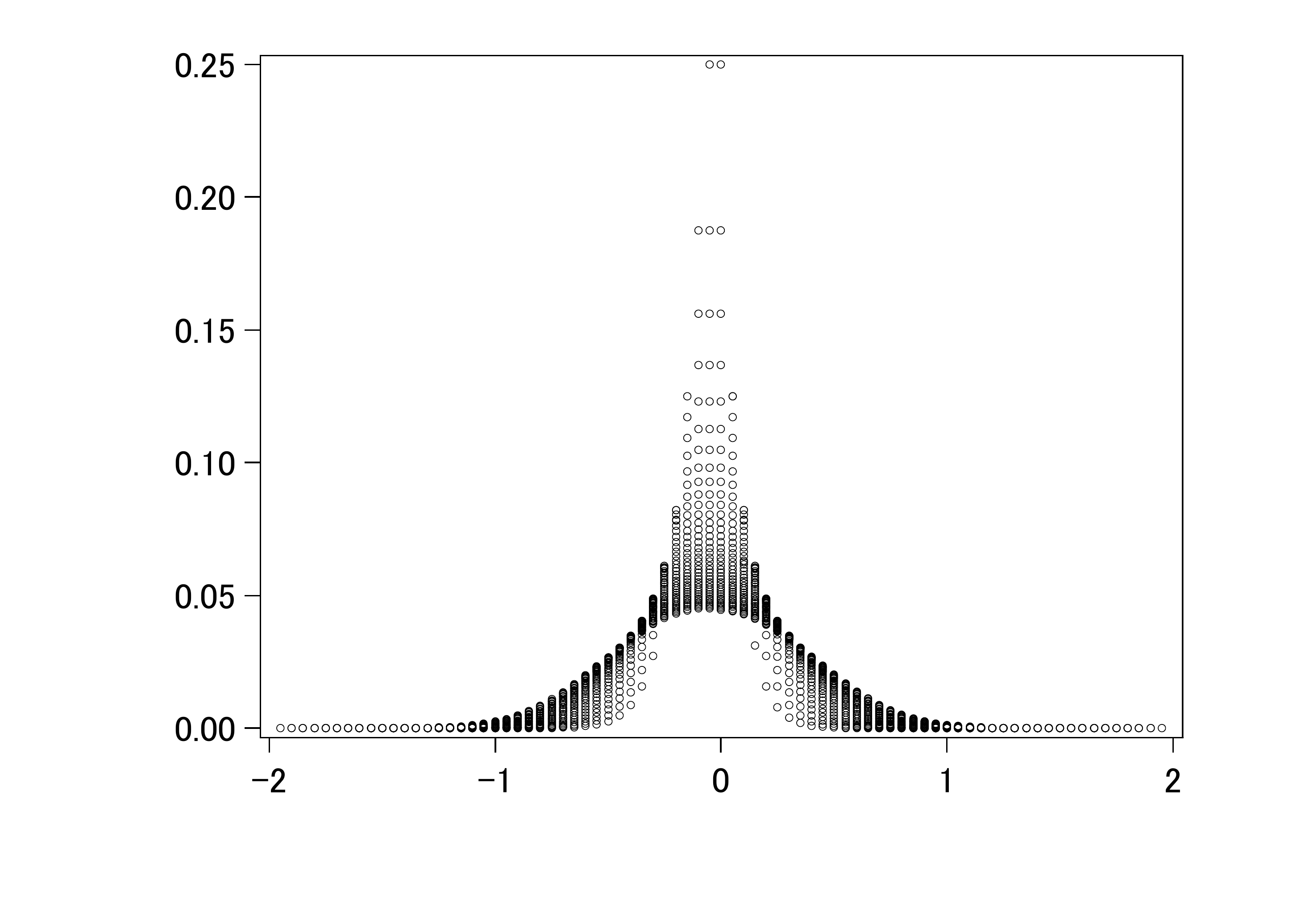}}}%
\end{minipage}
\begin{minipage}{100mm}
\subfigure[$n_1=n_2=50$ \label{figure1c}]{
\resizebox*{5cm}{!}{\includegraphics[height=5cm]{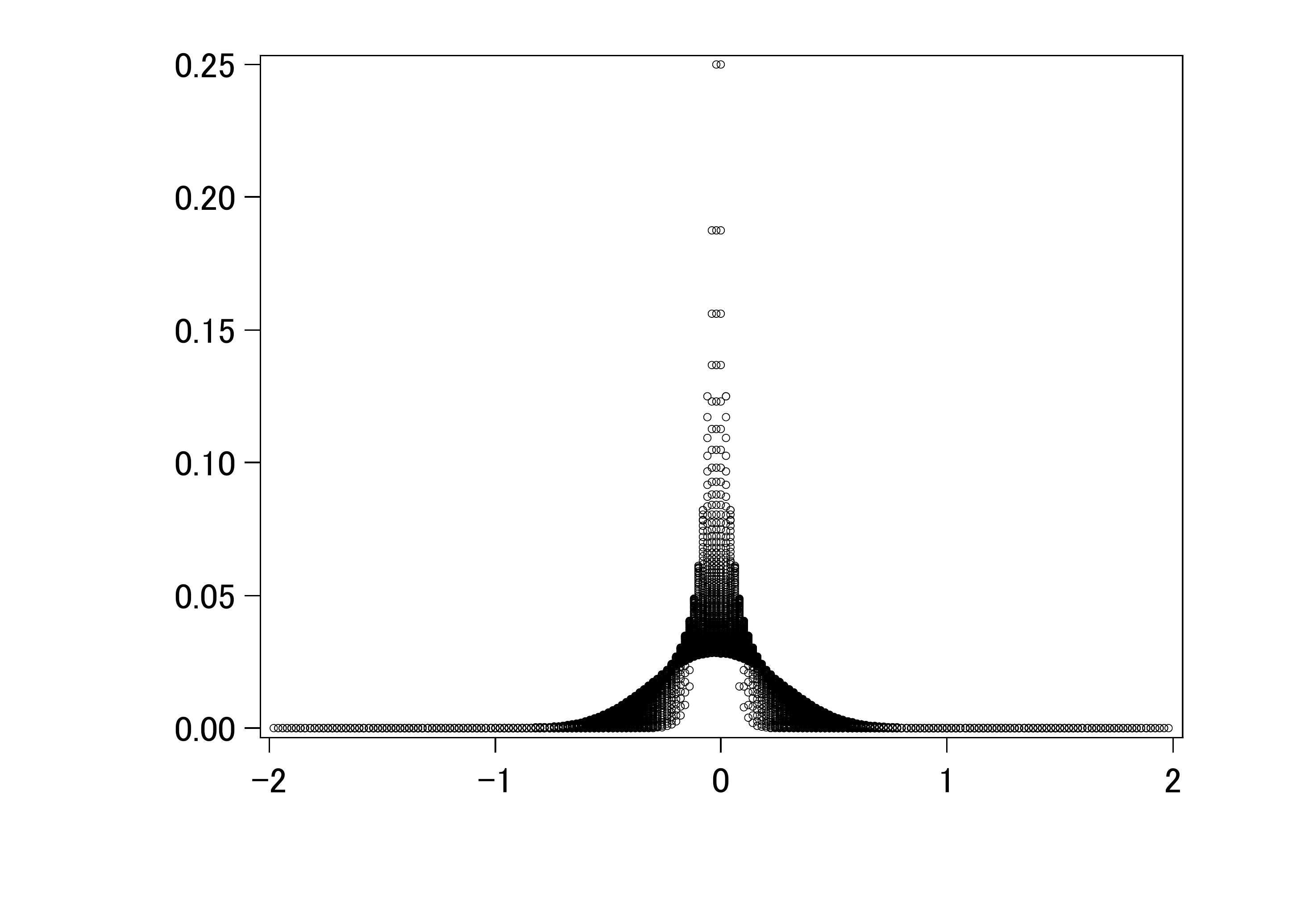}}}%
\subfigure[$n_1=n_2=100$ \label{figure1d}]{
\resizebox*{5cm}{!}{\includegraphics[height=5cm]{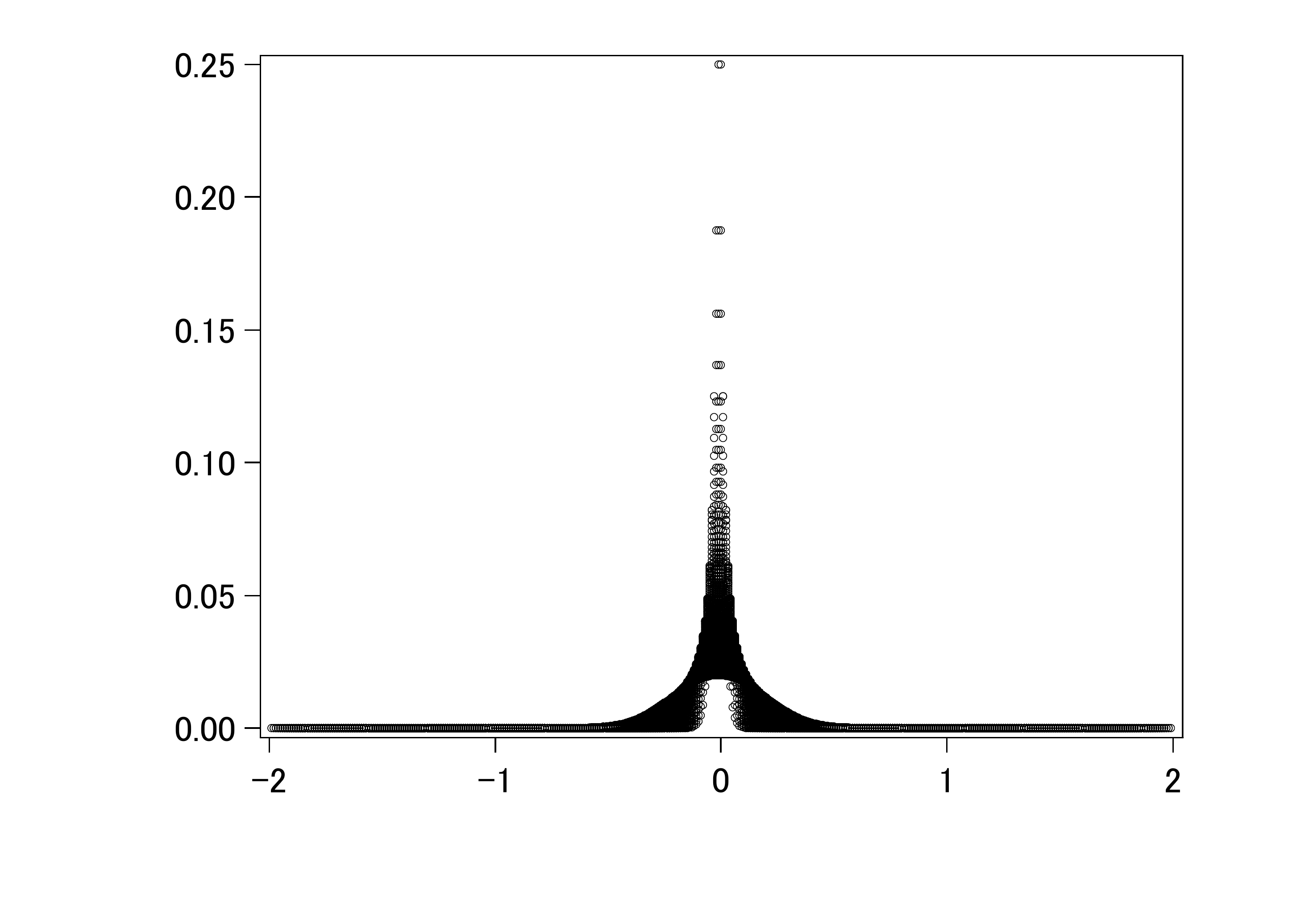}}}%
\caption{The comparison of $\theta - (1-p)$ and $\widehat{\lambda}_1 - \widehat{\lambda}_2 $ (vertical axis: $\theta - (1-p)$. horizontal axis: $\widehat{\lambda}_1 - \widehat{\lambda}_2 $ )}
\label{figure1}
\end{minipage}
\end{center}
\end{figure}

In Figure 2, the horizontal axis shows $1-p$, and the vertical axis shows $\theta$.\begin{figure}[h]
\begin{center}
\begin{minipage}{100mm}
\subfigure[$n_1=n_2=10$ \label{figure2a}]{
\resizebox*{5cm}{!}{\includegraphics[height=5cm]{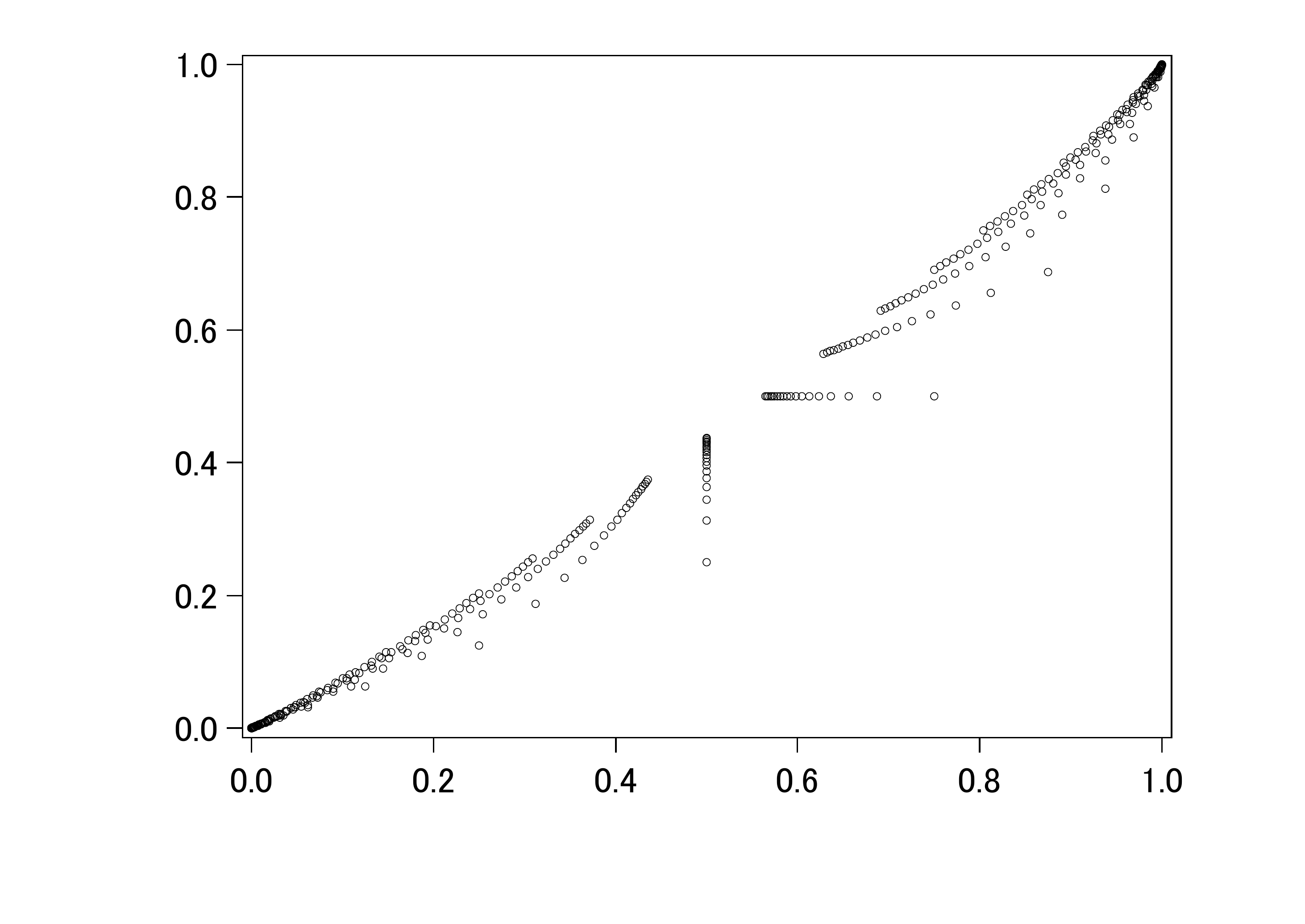}}}%
\subfigure[$n_1=n_2=20$ \label{figure2b}]{
\resizebox*{5cm}{!}{\includegraphics[height=5cm]{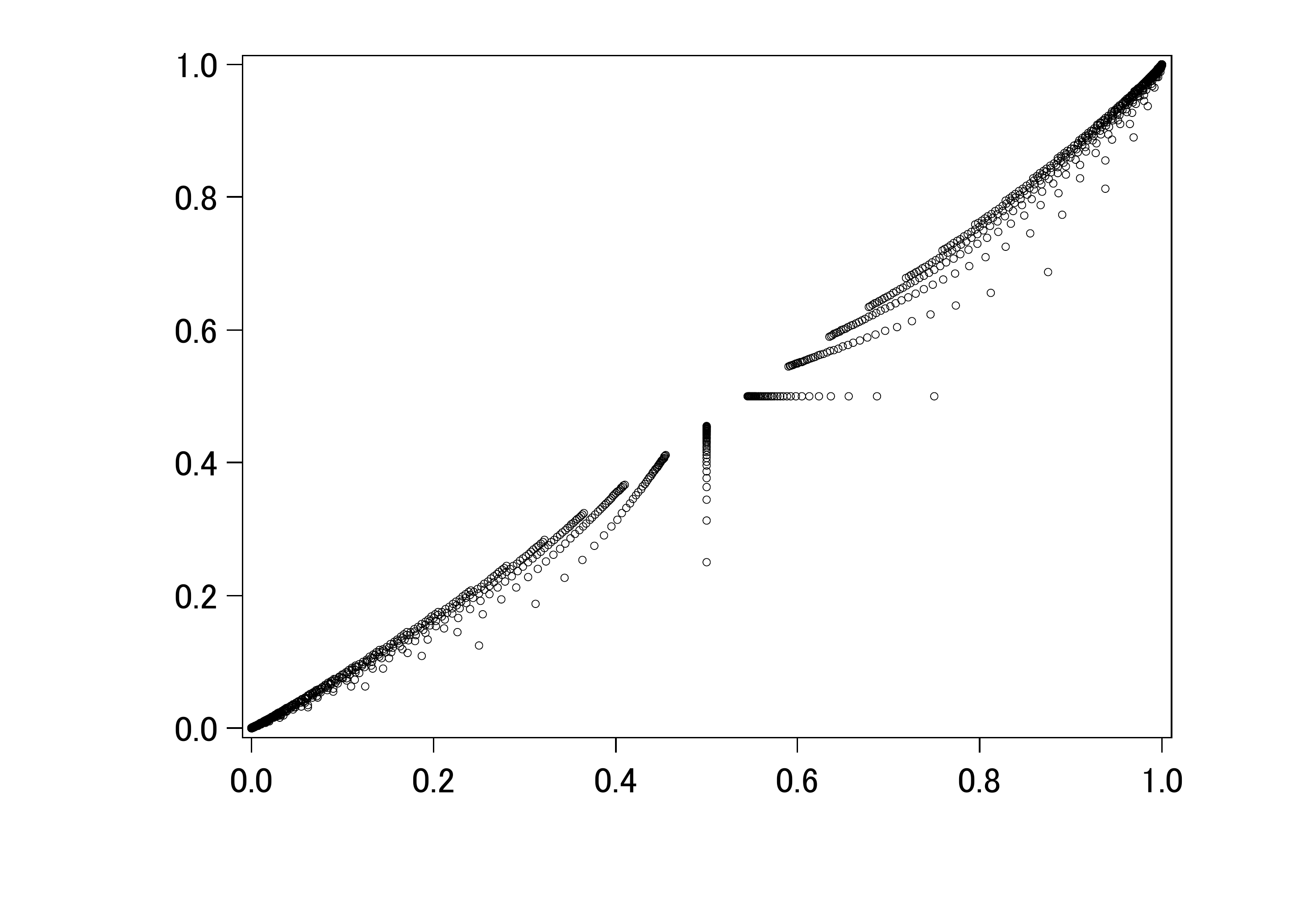}}}%
\end{minipage}
\begin{minipage}{100mm}
\subfigure[$n_1=n_2=50$ \label{figure2c}]{
\resizebox*{5cm}{!}{\includegraphics[height=5cm]{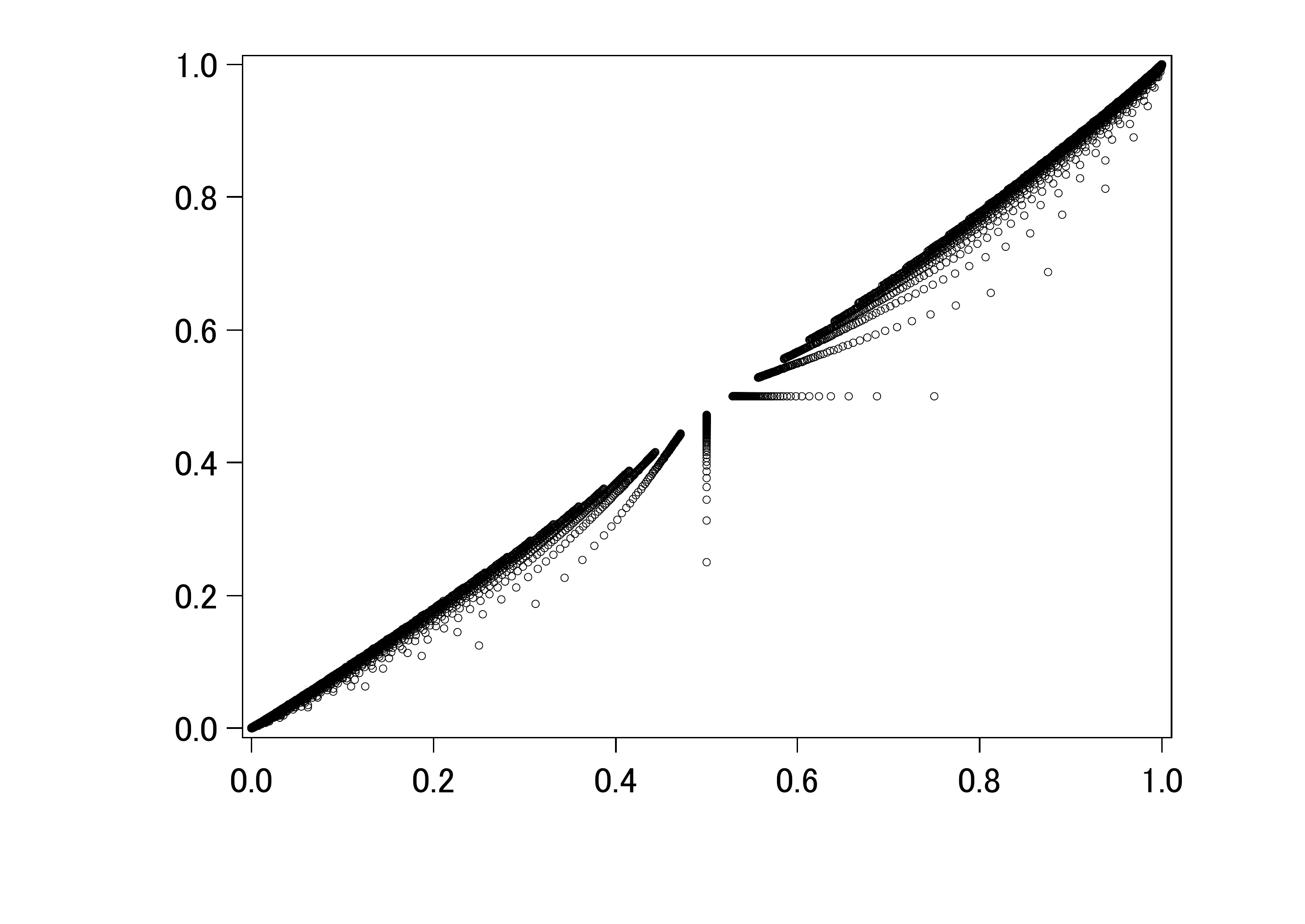}}}%
\subfigure[$n_1=n_2=100$ \label{figure2d}]{
\resizebox*{5cm}{!}{\includegraphics[height=5cm]{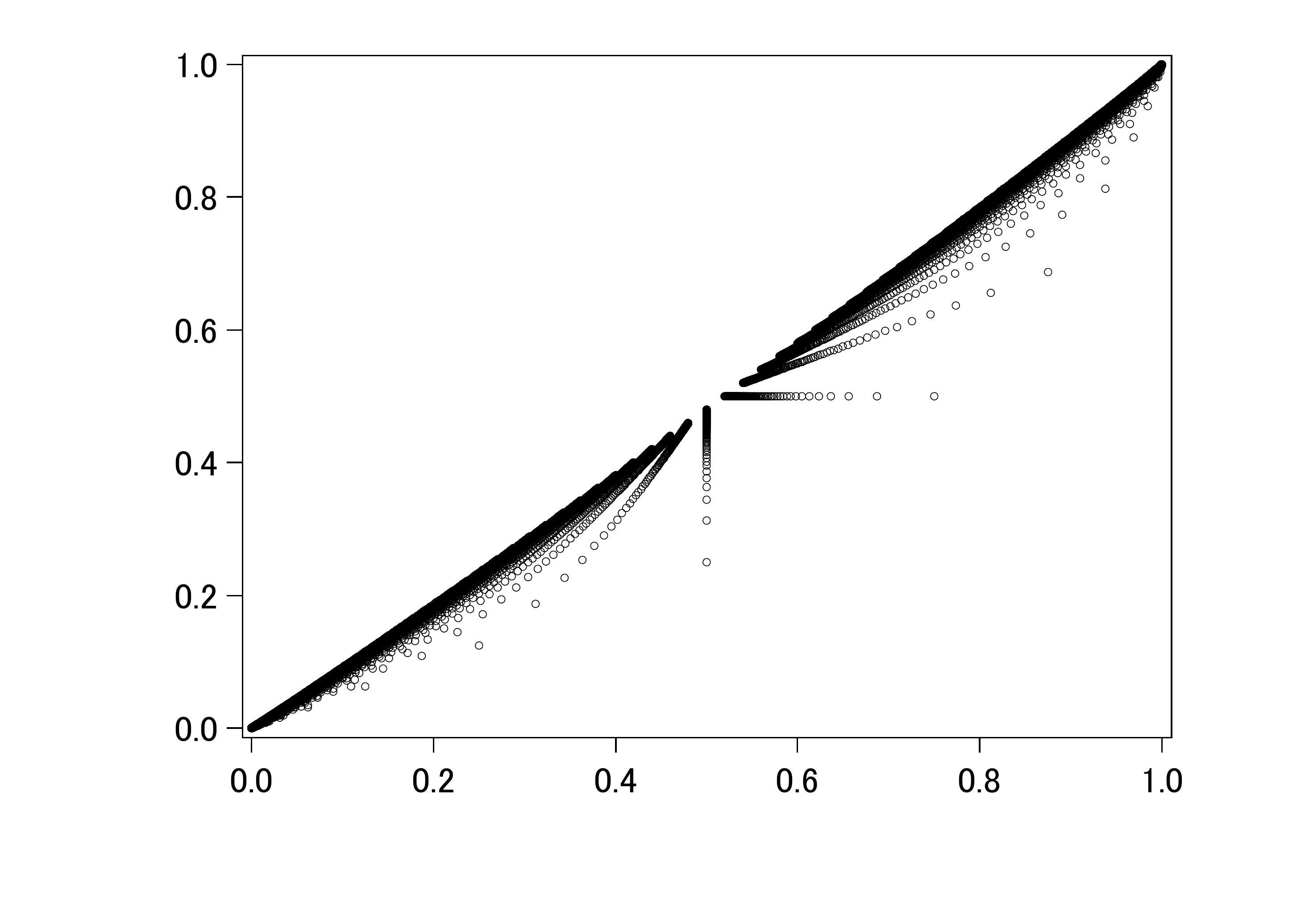}}}%
\caption{The comparison of $\theta$ given $X_1=k_1, X_2=k_2$ and $1-p$ given $X_1=k_1, X_2=k_2$ (vertical axis: $1-p$. horizontal axis: $\theta$)}
\label{figure2}
\end{minipage}
\end{center}
\end{figure}

In Figure 3, the horizontal axis shows $1-p$ given $X_1=k_1,X_2=k_2$, and the vertical axis shows $\theta$ given $X_1=k_1+1, X_2=k_2$. As theorem 2 states, $\theta$ always equals $1-p$.
\begin{figure}[h]
\begin{center}
\begin{minipage}{100mm}
\subfigure[$n_1=n_2=10$ \label{figure3a}]{
\resizebox*{5cm}{!}{\includegraphics[height=5cm]{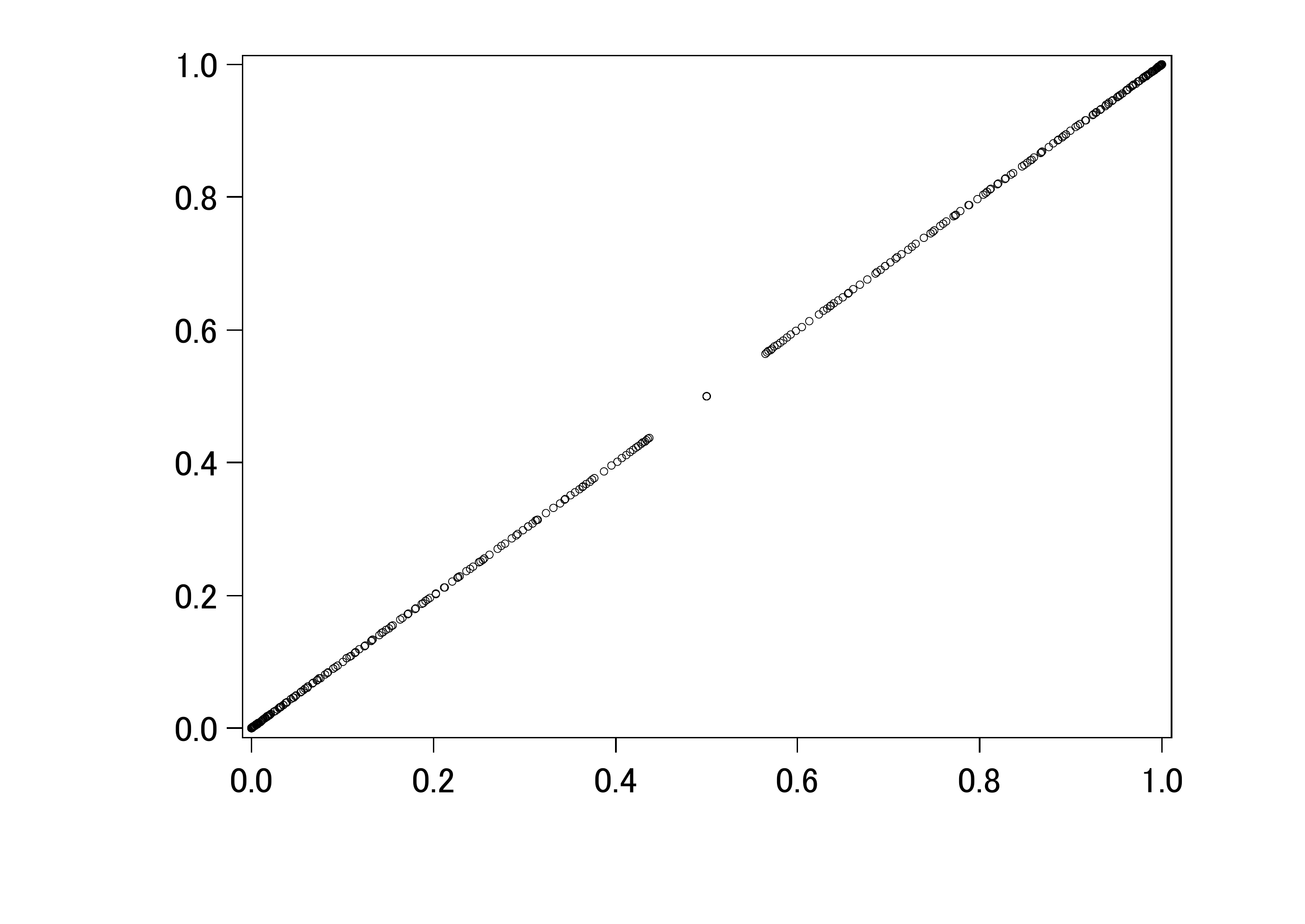}}}%
\subfigure[$n_1=n_2=20$ \label{figure3b}]{
\resizebox*{5cm}{!}{\includegraphics[height=5cm]{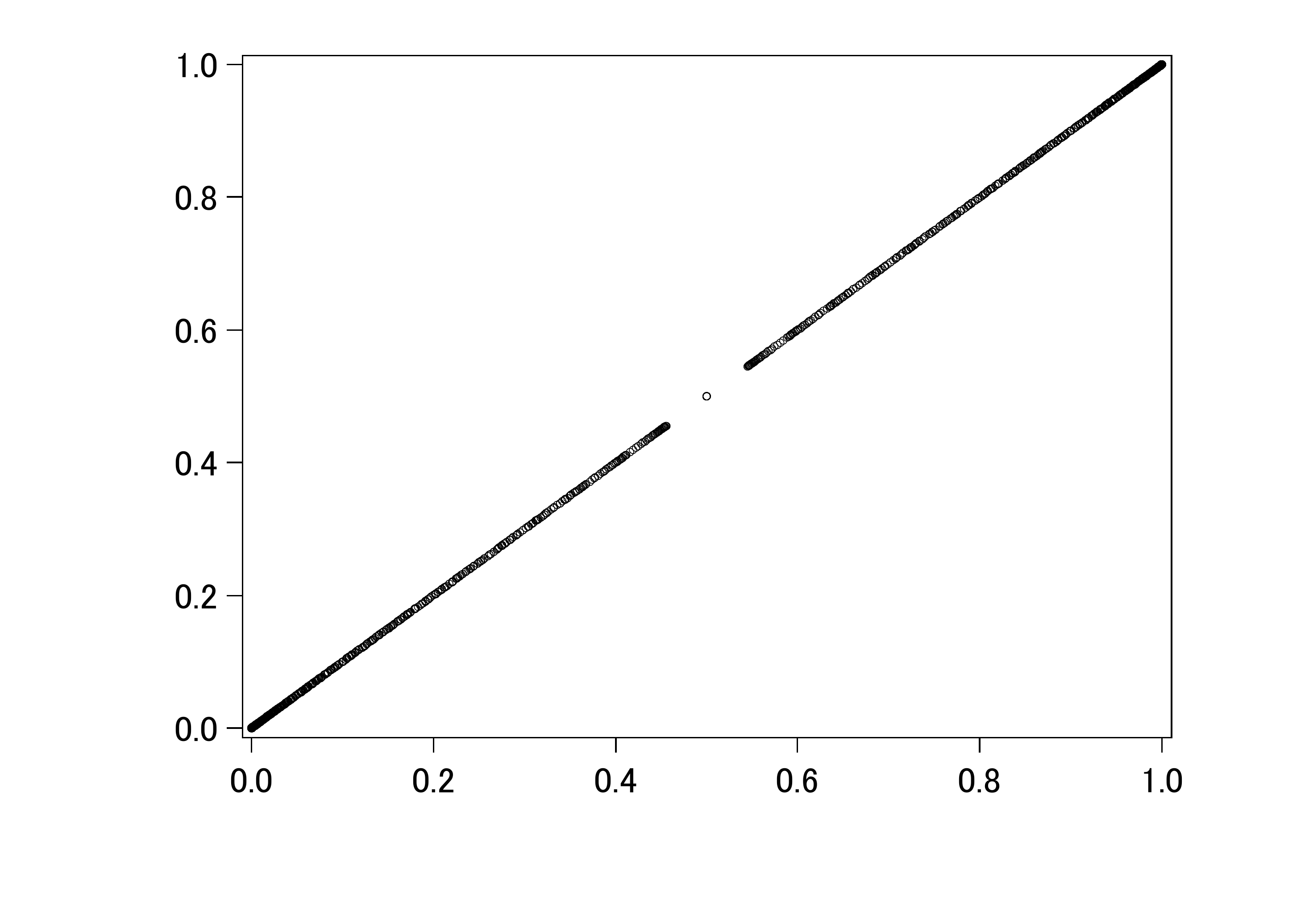}}}%
\end{minipage}
\begin{minipage}{100mm}
\subfigure[$n_1=n_2=50$ \label{figur3c}]{
\resizebox*{5cm}{!}{\includegraphics[height=5cm]{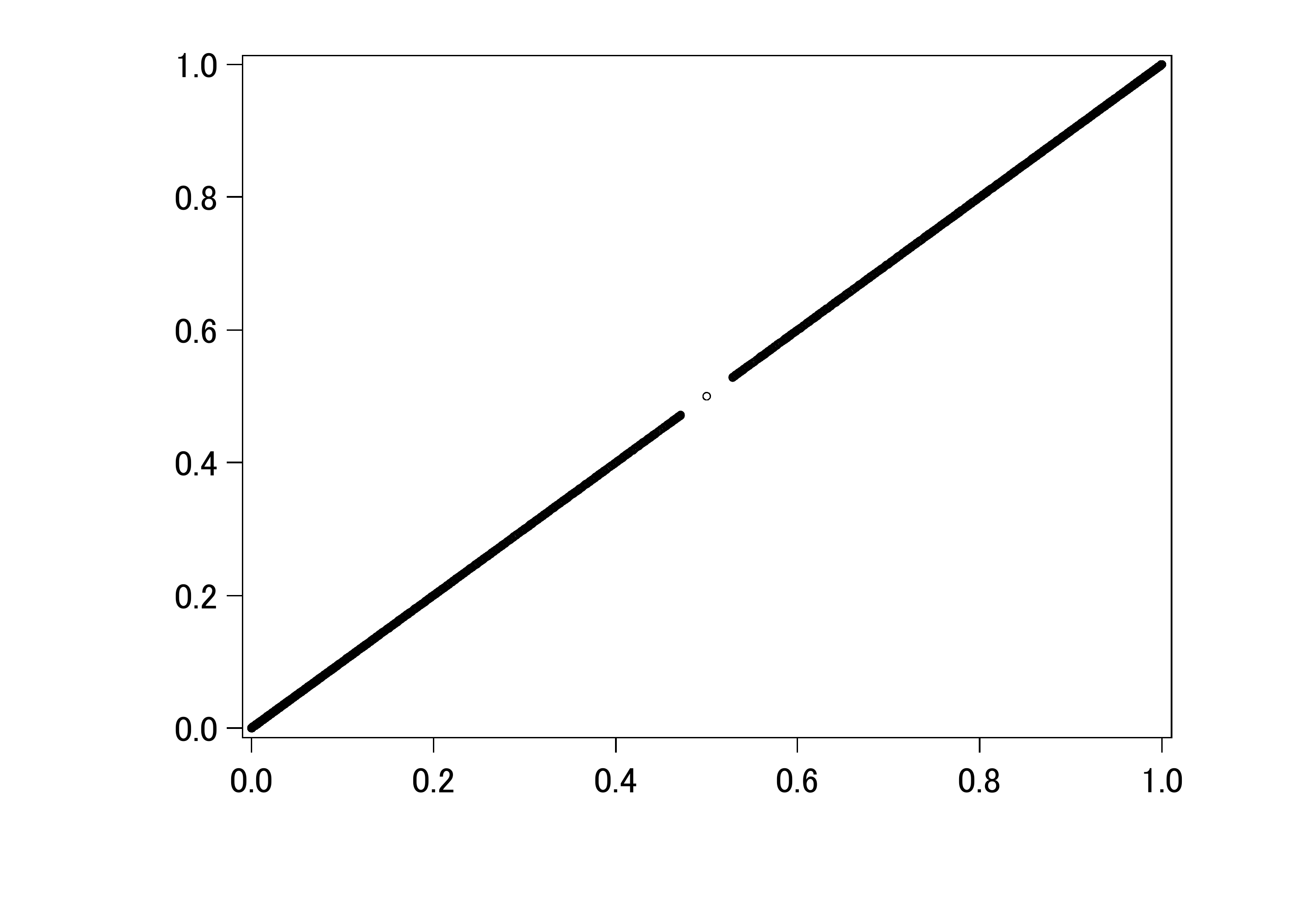}}}%
\subfigure[$n_1=n_2=100$ \label{figure3d}]{
\resizebox*{5cm}{!}{\includegraphics[height=5cm]{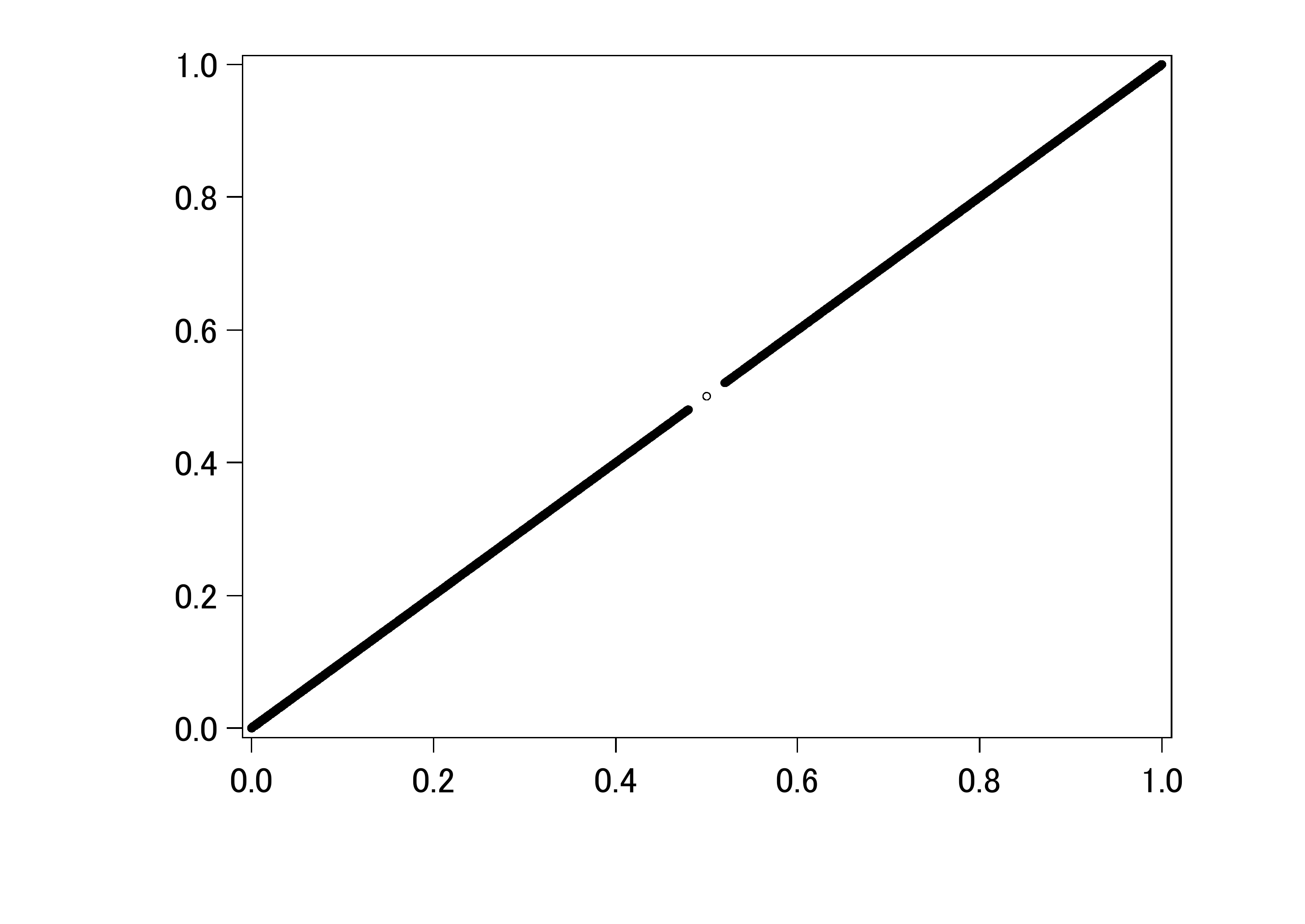}}}%
\caption{The comparison of $\theta$ given $X_1=k_1+1,X_2=k_2$ and $1-p$ given $X_1=k_1,X_2=k_2$ (vertical axis: $1-p$. horizontal axis: $\theta$)}
\label{figure3}
\end{minipage}
\end{center}
\end{figure}

\section{Generalization}
As a generalization for theorems 1 and 2, we consider the generalized version of the Bayesian index $\theta= P(\left. \lambda_1/\lambda_2 < c \, \right|\, X_1,X_2)$, and  
investigate the relationship between $\theta$ and the one-sided $p$-value of the conditional test with the null hypothesis $H_0: \lambda_1/\lambda_2 \geq c $ versus the  alternative $H_1: \lambda_1/\lambda_2 < c$. 
Let $\pi := \lambda_1/\lambda_2$. We consider the posterior of $\pi$ when the posterior of $\lambda_i$ is $Ga(a_i,b_i)$ for $i=1,2$. First, the joint density function of $(\lambda_1,\lambda_2)$ is 
\begin{eqnarray*}
 f(\lambda_1\, |\, a_1,b_1)\cdot f(\lambda_2\,|\, a_2,b_2) = \frac{b_1^{a_1}b_2^{a_2}}{\Gamma(a_1)\Gamma(a_2)} \lambda_1^{a_1-1} \lambda_2^{a_2-1} \exp( -(b_1\lambda_1 + b_2 \lambda_2) ).
\end{eqnarray*}
Next, let $\pi_1=\lambda_2$. Then, $\lambda_1=\pi \cdot \pi_1,\, \lambda_2=\pi_1$. Finally, the probability density function of the posterior distribution of $\pi$ is
\begin{eqnarray*}
f(\pi)&=& \int_0^{\infty} f(\pi \cdot \pi_1\,|\, a_1,b_1) \cdot f(\pi_1\,|\, a_2,b_2) \cdot
\left| \begin{array}{cc}
 \dfrac{\partial \lambda_1}{\partial \pi} &\dfrac{\partial \lambda_1}{\partial \pi_1}\\ 
 \dfrac{\partial \lambda_2}{\partial \pi} &\dfrac{\partial \lambda_2}{\partial \pi_1}\\  
 \end{array}
\right| 
d\pi_1\\
&=& \int_0^{\infty} 
\frac{b_1^{a_1}b_2^{a_2}}{\Gamma(a_1)\Gamma(a_2)} (\pi \cdot \pi_1)^{a_1-1} \pi_1^{a_2-1} \exp( -(b_1\pi \cdot \pi_1 + b_2 \pi_1) ) \cdot \pi_1 d\pi_1\\[5pt]
&=& \frac{b_1^{a_1}b_2^{a_2} \pi^{a_1-1}}{\Gamma(a_1)\Gamma(a_2)} \int_0^{\infty}
 \pi_1^{a_1+a_2-1} \exp(-\pi_1(b_1\pi + b_2)) d\pi_1\\[5pt]
&=& \frac{b_1^{a_1}b_2^{a_2} \pi^{a_1-1}}{\Gamma(a_1)\Gamma(a_2)} \cdot 
\frac{\Gamma(a_1+a_2)}{(b_1\pi +b_2)^{a_1+a_2}}
\\[5pt]
&=& \displaystyle \frac{1}{\pi B(a_1,a_2)} \left(
\frac{b_1 \pi}{ b_1 \pi + b_2}
\right)^{a_1} \left(
\frac{b_2}{ b_1 \pi + b_2}
\right)^{a_2}.
\end{eqnarray*}
Hence, the cumulative distribution function is
\begin{eqnarray}
F(x) &=& \int_0^x \displaystyle \frac{1}{\pi B(a_1,a_2)} \left(
\frac{b_1 \pi}{ b_1 \pi + b_2}
\right)^{a_1} \left(
\frac{b_2}{ b_1 \pi + b_2}
\right)^{a_2} d\pi
\\[5pt] \nonumber
&=&\frac{B_{\frac{b_1x}{b_1x + b_2}}(a_1,a_2)}{B(a_1,a_2)}\\\nonumber
&=& I_{\frac{b_1 x}{b_1x + b_2}}(a_1,a_2). \label{F2}
\end{eqnarray}
From this, we can obtain the expressions of the generalized version of the Bayesian index.
\proclaimit{Theorem\n 3.}{
If the posterior distribution of $\lambda_i$ is $Ga(a_i,b_i)$ with $a_i, b_i>0$ for $i=1,2$,
then, the Bayesian index $\theta=P( \lambda_1/\lambda_2 < c \,|\, X_1,X_2)$ has the following three expressions:
\begin{eqnarray}
\theta
&=& I_{\frac{b_1 c}{b_1 c+b_2}}(a_1, a_2) \label{p_100}\\
&=& F_{2a_1,2a_2}\left(\frac{b_1c/a_1}{b_2/a_2} \right) \nonumber\\
&=&1- \frac{1}{a_2 B(a_1,a_2) } \left(
\frac{b_2}{b_1c+b_2}
\right)^{a_2} \cdot {}_2 F_{1} \left( a_2, 1-a_1; 1+a_2; \frac{b_2}{b_1c+b_2} \right)  \nonumber
\end{eqnarray}
Additionally, if both $a_1$and $a_2$ are natural numbers, then, $\theta$ has the following two additional expressions:
\begin{eqnarray*}
\theta 
&=& \sum\limits_{r=0}^{a_2-1} {a_1+a_2-1 \choose r}
\left( \frac{b_2}{b_1c+b_2} \right)^r 
\left( \frac{b_1c}{b_1c+b_2} \right)^{a_1+a_2-1-r} \\[5pt]
&=& \sum\limits_{r=0}^{a_2-1} {a_1+r-1 \choose a_1 -1} \left( \frac{b_1c}{b_1c+b_2} \right)^{a_1} 
\left( \frac{b_2}{b_1c+b_2} \right)^{r}.
\end{eqnarray*}
}
{\sc Proof.}
(\ref{p_100}) can be shown as follows
\begin{eqnarray*} 
\theta&=&P\left(\left. \frac{\lambda_1}{\lambda_2} < c \,\right|\, X_1,X_2 \right)\\
&=&P\left(\left. \pi < c \,\right|\, X_1,X_2 \right) \nonumber \\
&=&F(c) \nonumber \\
&=& I_{\frac{b_1 c}{b_1c + b_2}}(a_1,a_2)  \hspace{5mm}(\because (\ref{F2})).
\end{eqnarray*}
The remainder of the proof is almost the same as that of theorem 1.
\begin{flushright}
$\square$
\end{flushright}

On the other hand, the one-sided $p$-value of the conditional test with $H_0: \lambda_1/\lambda_2 \geq c$ versus $H_1:\lambda_1 / \lambda_2 < c$ 
(Przyborowski and Wilenski, 1940; Krishnamoorthy and Thomson, 2004) is defined as
\begin{eqnarray*}
p&=& P(X_1 \leq k_1 \, |\, X_1+X_2=k_1+k_2, \lambda_1/\lambda_2 = c)\\
&=& \sum_{r=0}^{k_1} {k_1+k_2 \choose r} \left( \frac{n_1c}{n_1 c+n_2} \right)^r 
\left(
\frac{n_2}{n_1c + n_2}
\right)^{k_1+k_2-r}.
\end{eqnarray*}
Then, we can obtain the following lemma.
\proclaimit{ Lemma\n 2.}{If $k_2>0$, then the one-sided $p$-value of the conditional test with $H_0:\lambda_1/\lambda_2 \geq c$ vs. $H_1:\lambda_1/\lambda_2 <c$ has the following expressions:
\begin{eqnarray*}
p &=& I_{\frac{n_2}{n_1c+n_2}}(k_2,k_1+1)\\
&=&F_{2k_2,2(k_1+1)}\left( \frac{n_2/k_2}{n_1c /(k_1+1)} \right)\\
&=& \frac{1}{k_2B(k_2,k_1+1)} \left( \frac{n_2}{n_1c+n_2} \right)^{k_2} \cdot {}_2F_1
\left(
k_2,-k_1;1+k_2; \frac{n_2}{n_1c+n_2}
\right).
\end{eqnarray*}
If $k_2=0$, then $p=1$.
}
{\sc Proof.}
The proof is almost the same as lemma 1.
\begin{flushright}
$\square$
\end{flushright}

Finally, we can obtain the generalization of theorem 2.
\proclaimit{Theorem\n 4.}{
If $k_2>0$, then between $\theta=P(\lambda_1/\lambda_2 < c \,|\, X_1,X_2)$ given $X_1=k_1+1$,  $X_2=k_2$ and the one-sided $p$-value of the conditional test with $H_0:\lambda_1 / \lambda_2 \geq c$ vs. $H_1:\lambda_1 /\lambda_2< c$ given $X_1=k_1, X_2=k_2$, the following relation holds
\begin{eqnarray*}
\lim\limits_{(\alpha_1,\alpha_2,\beta_1,\beta_2)\to (0,0,0,0)} \theta = 1- p.
\end{eqnarray*}
}

{\sc Proof.}
From lemma 2, the $p$-value given $X_1=k_1, X_2=k_2$ is
\begin{eqnarray*}
p&=&  I_{\frac{n_2}{n_1c+n_2}}(k_2,k_1+1).
\end{eqnarray*}
Therefore, from the relation $I_z(a,b)=1-I_{1-z}(b,a)$,
\begin{eqnarray}
1-p&=&  I_{\frac{n_1c}{n_1c+n_2}}(k_1+1,k_2). \label{p_value101}
\end{eqnarray}
On the other hand, from (\ref{p_100}), the Bayesian index given $X_1=k_1+, X_2=k_2$ can be expressed as
\begin{eqnarray}
\theta &=& I_{\frac{b_1c}{b_1c+b_2}}(a_1,a_2) \nonumber\\
&=&I_{\frac{(\beta_1+n_1)c}{(\beta_1+n_1)c+(\beta_2+n_2)}}(\alpha_1+k_1+1, \alpha_2+k_2). \label{theta102}\\
&&(\because a_1=\alpha_1+k_1+1, a_2=\alpha_2+k_2, b_1=\beta_1+n_1, b_2=\beta_2+n_2) \nonumber
\end{eqnarray}
Finally, from (\ref{theta102}) and (\ref{p_value101}), 
\begin{eqnarray*}
\lim_{(\alpha_1,\alpha_2,\beta_1,\beta_2) \to (0,0,0,0)} \theta = 1-p.
\end{eqnarray*}
holds. We have completed the proof of theorem 4.
\begin{flushright}
$\square$
\end{flushright}

\section{Application}
In this section, we apply the Bayesian index to real epidemiology and clinical trial data and compare it to the one-sided $p$-value of the conditional test.
\proclaimit{Example\n 4.}{Breast cancer study;}
Table 1 shows the result of a breast cancer study reported in Rothman and Greenland (2008). The rates of breast cancer between two groups of women are compared. One group is composed of the women with tuberculosis who are repeatedly exposed to multiple x-ray fluoroscopies and the other group is composed of unexposed women with tuberculosis. 
\begin{table}[h]
\centering
\caption{Breast cancer study data }
\begin{tabular}{ccc}
\hline 
&cases of breast cancer $(X_i)$ & person-years at risk $(n_i)$\\ \hline
Received x-ray fluoroscopy
$(i=1)$& 41 & 28,010
\\
Control $(i=2)$& 15 & 19,017 
\\ \hline
\end{tabular}
\end{table}

Let $X_1,X_2$ be the independent Poisson processes indicating the numbers of cases of breast cancer, and $n_1,n_2$ be person-years at risk. Here, we suppose $X_i \stackrel{ind}{\sim} Po(n_i \lambda_i)$ for $i=1,2$. From Table 1, $X_1=41, n_1=28,010$ and $X_2=15, n_2=19,017$. First, we consider the conditional test with the null hypothesis 
$H_0:\lambda_1 \leq \lambda_2 $ versus the alternative $H_1:\lambda_1 > \lambda_2 $ and the Bayesian index $\theta=P(\lambda_1 > \lambda_2\,|X_1,X_2,n_1,n_2)$ with the  non-informative and Jeffrey's priors. Table 2 shows the results. Here, $1-p=0.976$. Hence, $\theta-(1-p)=0.009$ and $0.007$ with the  non-informative and Jeffrey's priors, respectively.
\begin{table}[h]
\centering
\caption{$p$-value with $H_0: \lambda_1 \leq \lambda_2$ vs. $H_1: \lambda_1 > \lambda_2$ and Bayesian index $P(\lambda_1 > \lambda_2 \,|\, X_1,X_2)$ for the breast cancer study data}
\begin{tabular}{ccc}
\hline 
$p$-value with & \multicolumn{2}{c}{Bayesian index $P(\lambda_1 > \lambda_2 \,|\, X_1,X_2)$}\\ \cline{2-3}
$H_0: \lambda_1 \leq \lambda_2$ vs. $H_1: \lambda_1 > \lambda_2$ & non-informative prior &Jeffrey's prior \\\hline
0.024 & 0.985 & 0.983
\\ \hline
\end{tabular}
\end{table}

Next, as in Gu et al. (2008), we consider the test with the null hypothesis 
$H_0:\lambda_1 / \lambda_2 \leq 1.5$ versus the alternative $H_1:\lambda_1 / \lambda_2 > 1.5 $ and the Bayesian index $P(\lambda_1/ \lambda_2 > 1.5 \,|\, X_1,X_2)$ with the  non-informative and Jeffrey's priors. Table 3 shows the results. In this case, $1-p=0.709$. Hence, $\theta-(1-p)=0.067$ and $0.048$ with the non-informative and Jeffrey's priors, respectively.
\begin{table}[h]
\centering
\caption{$p$-value with $H_0: \lambda_1 / \lambda_2 \leq 1.5 $ vs. $H_1: \lambda_1/\lambda_2 > 1.5$ and Bayesian index $P(\lambda_1/\lambda_2 >1.5 \,|\, X_1,X_2)$ for the breast cancer study data}
\begin{tabular}{ccc}
\hline 
$p$-value with & \multicolumn{2}{c}{Bayesian index $P(\lambda_1/\lambda_2 >1.5 \,|\, X_1,X_2)$}\\ \cline{2-3}
$H_0: \lambda_1 / \lambda_2 \leq 1.5 $ vs. $H_1: \lambda_1/\lambda_2 > 1.5$ & non-informative prior &Jeffrey's prior \\\hline
0.291 & 0.776 & 0.757
\\ \hline
\end{tabular}
\end{table}
\proclaimit{Example\n 5.}{Hypertension trials;}
Table 4 shows the result of the two selected hypertension clinical trials in Table II in Arends et al. (2000). We assume that the trial 1 is of interest and we utilize the trial 2 data to specify the conditional power priors described in section \ref{eg}
Let $X_1,X_2$ be the independent Poisson process indicating the number of deaths, and $n_1,n_2$ be the number of the person-year in trial 1, and let $x^0_1,x^0_2$ be the independent Poisson process indicating the number of deaths, and $m_1,m_2$ be the number of the person-year in trial 2. Here we suppose $X_i \sim Po(n_i \lambda_i),\, x_i^0 \sim Po(m_i \lambda_i)$, and $X_1,X_2,x_1^0,x_2^0$ are independent. From Table 4, $X_1=54, n_1=5,635,\, X_2=70, n_2=5,600,\,x_1^0=47, m_1=5,135,\, x_2^0=63, m_2=4,960$.
\begin{table}[h]
\centering
\caption{Hypertension trials data }
\begin{tabular}{ccccccc}
\hline 
&\multicolumn{2}{c}{Treatment group} && \multicolumn{2}{c}{Control group}\\\cline{2-3}\cline{5-6}
& death & number of person-year & & death & number of person-year\\ \hline 
Trial 1& 54 & 5,635 & & 70 & 5,600\\
Trial 2& 47 & 5,135 & & 63 & 4,960\\
\hline
\end{tabular}
\end{table}

We consider the test with the null hypothesis 
$H_0:\lambda_1 \geq \lambda_2 $ versus the alternative $H_1:\lambda_1 < \lambda_2 $ and the Bayesian index $\theta=P(\lambda_1 < \lambda_2 \,|\, X_1,X_2)$ with the non-informative prior and the conditional power priors. For the conditional power prior, we assume $a_1=a_2(=:a)$ and take $a=0.1, 0.5$, and $1.0$. Table 5 shows the result. Here $1-p=0.917$ and $\theta$ with the non-informative prior is 0.930.  Additionally, $\theta$ with the conditional power priors are greater than that with the non-informative prior. Moreover, when $a$ increases, $\theta$ also increases. 
\begin{table}[h]
\centering
\caption{$p$-value and Bayesian index for the hypertension trials data ($a=a_1=a_2$)}
\begin{tabular}{cccccc}
\hline 
$p$-value with  &\multicolumn{5}{c}{Bayesian index $P(\lambda_1 < \lambda_2 \,|\, X_1,X_2)$} \\ \cline{2-6} 
$H_0:\lambda_1 \geq \lambda_2$ vs. $H_1: \lambda_1 < \lambda_2$  &non-informative prior&  & \multicolumn{3}{c}{conditional power prior}\\ \cline{2-2} \cline{4-6}
&  && $a=0.1$ & $a=0.5$ & $a=1.0$
\\
\hline
0.083 & 0.930 & & 0.942 & 0.971 & 0.988 \\ \hline
\end{tabular}
\end{table}

Here, suppose that $\theta \geq 0.95$ is effective. Then, with the non-informative prior, effectiveness is similar to $p<0.05$ because $\theta$ is similar to $1-p$. On the other hand, with the conditional power prior with suitable historical data, effectiveness is more easily satisfied than $p<0.05$. 

\section{Conclusion}
In this paper, we provided the cumulative distribution function expressions of the Bayesian index $\theta=P(\lambda_1<\lambda_2 \,|\, X_1,X_2)$ for the Poisson parameters, which can be more easily calculated than the hypergeometric series expression in Kawasaki and Miyaoka (2012b). Next, we showed the relationship between the Bayesian index with the non-informative prior and the one-sided $p$-value of the conditional test with $H_0:\lambda_1 \geq \lambda_2$ versus $H_1:\lambda_1 < \lambda_2$. This relationship can be considered as the Poisson distribution counterpart of the relationship between the Bayesian index for binomial proportions and the one-sided $p$-value of the Fisher's exact test in Kawasaki et al. (2014). Additionally, we generalized the Bayesian index to $\theta=P(\lambda_1/\lambda_2 < c\,|\, X_1,X_2)$, expressed it using the cumulative distribution functions and hypergeometric series, and investigated the relationship between $\theta$ and the one-sided $p$-value of the conditional test with $H_0: \lambda_1/\lambda_2 \geq c$ versus $H_1:\lambda_1/\lambda_2 < c$. By the analysis of real data, we showed that the Bayesian index with the non-informative prior is similar to  $1-p$ of the conditional test and the Bayesian index with the conditional power prior with suitable historical data can potentially
improve the efficiency of inference.

%
%
%
%
\references
\renewcommand{\labelenumi}{[\arabic{enumi}]}
\renewcommand{\makelabel}{\rm}
\setcounter{enumi}{0}
\begin{enumerate}
\item Abramowitz, M., and Stegun, I. A. (1964). {\it Handbook of mathematical functions: with formulas, graphs, and mathematical tables (No. 55)}. Courier Corporation.
\item Altham, P. M. (1969). Exact Bayesian analysis of a $2 \times 2$ contingency table, and Fisher's" exact" significance test. {\it Journal of the Royal Statistical Society. Series B (Methodological)}, \textbf{31}(2) ,261-269.
\item Arends, L. R., Hoes, A. W., Lubsen, J., Grobbee, D. E., and Stijnen, T. (2000). Baseline risk as predictor of treatment benefit: three clinical meta-re-analyses. \textit{Statistics in Medicine}, \textbf{19}(24), 3497-3518.
\item Gu, K., Ng, H. K. T., Tang, M. L., and Schucany, W. R. (2008). Testing the ratio of two poisson rates. {\it Biometrical Journal}, {\bf 50}(2), 283-298.
\item Howard, J. V. (1998). The $2\times 2$ table: A discussion from a Bayesian viewpoint. {\it Statistical Science},\textbf{13}(4) 351-367.
\item Ibrahim, J. G., and Chen, M. H. (2000). Power prior distributions for regression models. {\it Statistical Science}, \textbf{15}(1), 46-60.
\item Jeffreys, H. (1946). An invariant form for the prior probability in estimation problems. \textit{Proceedings of the Royal Society of London} (Ser A), \textbf{186},  453-461.
\item Kawasaki, Y., and Miyaoka, E. (2012a). A Bayesian inference of $P (\pi_1> \pi_2)$ for two proportions. {\it Journal of Biopharmaceutical Statistics}, {\bf 22}(3), 425-437.
\item Kawasaki, Y., and Miyaoka, E. (2012b). A Bayesian inference of $P(\lambda_1 < \lambda_2)$ for two Poisson parameters. {\it Journal of Applied Statistics}, {\bf 39} (10), 2141-2152.
\item Kawasaki, Y., and Miyaoka, E. (2014). Comparison of three calculation methods for a Bayesian inference of two Poisson parameters. \textit{Journal of Modern Applied Statistical Methods}, \textbf{13}(1), 397-409.
\item Kawasaki, Y., Shimokawa, A., and Miyaoka, E. (2013). Comparison of three calculation methods for a Bayesian inference of $P (\pi_1>\pi_2)$. \textit{Journal of Modern Applied Statistical Methods}, \textbf{12}(2), 256-268.
\item Kawasaki, Y., Shimokawa, A., and Miyaoka., E. (2014). On the Bayesian index of superiority and the $P$-value of the Fisher exact test for binomial proportions. {\it Journal of the Japan Statistical Society}. {\bf 44}(1), 73-81.
\item Krishnamoorthy, K., and Thomson, J. (2004). A more powerful test for comparing two Poisson means. {\it Journal of Statistical Planning and Inference}, {\bf 119}(1), 23-35.
\item Przyborowski, J., and Wilenski, H. (1940). Homogeneity of results in testing samples from Poisson series: With an application to testing clover seed for dodder. {\it Biometrika}, \textbf{31}(3), 313-323.
\item Rothman, K. J., Greenland, S., and Lash, T. L. (2008). \textit{Modern epidemiology. 3rd}. Philadephia: Lippincott Williams $\&$ Wilkins.
\item Zwillinger, D. (Ed.). (2014). {\it Table of integrals, series, and products}. Elsevier.
\end{enumerate}
\end{document}